\documentclass[twocolumn]{aa}
\usepackage{graphicx}
\usepackage{txfonts}
\usepackage{natbib}
\begin{document}
   \title{The 1 -- 50 keV spectral and timing analysis of IGR J18027-2016:
         an eclipsing, high mass X-ray binary.}

   \author{A.B. Hill,
          \inst{1}
	  R. Walter\inst{2},
	  C. Knigge\inst{1},
	  A. Bazzano\inst{3}
	  G. B\'{e}langer\inst{4}, A.J. Bird\inst{1}, A.J. Dean\inst{1}, J.~L. Galache\inst{1},
          A. Malizia\inst{5}, M. Renaud\inst{4}, J. Stephen\inst{5} \and P. Ubertini\inst{3} 
          }

   \authorrunning{A.~B. Hill et al.}
   \titlerunning{The 1--50 keV spectral and timing analysis of IGR J18027-2016} 

   \offprints{A.~B. Hill, email: abh@astro.soton.ac.uk}

   \institute{School of Physics and Astronomy, University of Southampton, Highfield, SO17 1BJ, UK.
	      \and
	      Geneva Observatory, INTEGRAL Science Data Centre, Chemin d'Ecogia 16, 1291 Versoix, Switzerland.
              \and
	      IASF-Rm, INAF, Via Fosso del Cavaliere 100, I-00133 Rome, Italy.
	      \and
	      CEA-Saclay, DAPNIA/Service d'Astrophysique, F-91191 Gif sur Yvette Cedex, France.             
	      \and
	      IASF-Bo, INAF, Via Gobetti 101, I-40129 Bologna, Italy.\\
}      

   \date{Received 25 February 2005 / Accepted 24 April 2005}

   \abstract{We report the association of the INTEGRAL source
   IGR J18027-2016 with the BeppoSAX source SAX J1802.7-2017.  IGR
   J18027-2016 is seen to be a weak, persistent source by the
   IBIS/ISGRI instrument on board INTEGRAL with an average source
   count rate of 0.55 counts s$^{-1}$ ($\sim$6.1 mCrab) in
   the 20-40 keV band.  Timing analysis performed on the ISGRI
   data identifies an orbital period of 4.5696 $\pm$ 0.0009 days
   and gives an ephemeris of mid-eclipse as, $T_{mid}$ = 52931.37 $\pm$ 0.04 MJD.
   Re-analysis of archival BeppoSAX data has
   provided a mass function for the donor star, $f(m) = 16 \pm 1 M_{\sun}$ and a
   projected semimajor axis of  $a_{x}\sin{i} = 68 \pm 1$ lt-s.
   We conclude that the donor is an OB-supergiant with a mass of
   18.8--29.3 M$_{\sun}$ and a radius of 15.0--23.4 R$_{\sun}$.
   Spectra obtained by XMM-Newton and ISGRI indicate a high
   hydrogen column density of N$_{H}$ = 6.8 $\times$ 10$^{22}$ cm$^{-2}$, which
   suggests intrinsic absorption.  The source appears to be a
   high mass X-ray binary with the neutron star emitting X-rays
   through wind-fed accretion while in an eclipsing orbit
   around an OB-supergiant.

   }

   \maketitle
%

\section{Introduction}

High mass X-ray binaries (HMXBs) were discovered as bright X-ray
sources in the 1970s.  The systems comprise a compact object
orbiting a massive OB class star.  The compact object can be either a
neutron star or black hole and is a strong X-ray emitter via the
accretion of matter from the OB companion.
HMXBs can be split into two individual classes: ({\it i}) Be/X-ray
binaries; ({\it ii}) Supergiant X-ray binaries \citep{Corbet}.  The
majority of HMXBs belong to the first class ($\sim$80$\%$)
\citep{hmxb_percent}.  In Be systems, the compact object is a 
neutron star and is typically in a wide, moderately eccentric orbit and
spends little time in close proximity to the dense circumstellar disk
surrounding the Be companion, \citep{HMXB_Coe, HMXB_Liu}.  As a result,
they are often X-ray transients as the accretion principally occurs
when the compact object passes through the Be-star disk. 

In the supergiant systems, the compact object is in orbit around an
OB-supergiant.  Accretion can be powered solely by the stellar wind
from the supergiant and/or can be through Roche-lobe overflow.  Accretion
from the stellar wind results in an X-ray luminosity of
10$^{35}$--10$^{36}$ erg s$^{-1}$.  When the star fills its Roche lobe
a much higher X-ray luminosity is achieved, $\sim$10$^{38}$ erg
s$^{-1}$ \citep{hmxb_percent}.  The orbital periods of the supergiant
systems are typically shorter (1.4 -- 41.5 days) than the Be-star
systems (12.7 -- 262 days) and the orbits typically more circular.

HMXBs are young systems, and the compact object is often a highly
magnetized neutron star.  Hence, many systems are seen to host an
X-ray pulsar.  The spin period of the pulsar and the orbital period of
the system correlate well allowing the differentiation between
OB-supergiant and Be-star systems and between underfilled and filled
Roche-lobe supergiant systems \citep{Corbet}.

Since its launch in 2002, INTEGRAL (the INTErnational Gamma-Ray
Astrophysics Laboratory) has been performing a regular survey of the
Galactic plane and a deep exposure of the Galactic Centre as part of
its Core Programme \citep{integral}.  In the course of these
observations a number of new unidentified sources emitting in the soft
$\gamma$-ray region have been discovered.  The first IBIS/ISGRI soft
$\gamma$-ray galactic plane survey catalogue \citep{ibissurvey}
reports the discovery of 28 objects of unknown classification
exhibiting persistent hard X-ray emission.  One of the new
unidentified sources observed in the first year of operations of
INTEGRAL was IGR J18027-2016.  This new source has been independently
reported in both core programme GCDE (Galactic Centre Deep Exposure)
observations \citep{ibissurvey} and guest observer data \citep{russian1}.

IGR J18027-2016 is spatially associated with a recently identified
X-ray pulsar SAX J1802.7-2017 \citep{sax}.  \citet{luto_gc} report
that the 18--60 keV spectrum of IGR J18027-2016 can be described by a
power law with an exponential cut-off at high energies of E$_{cut}$ =
18 keV.  We report here the results of the
timing analysis of IGR J18027-2016 using INTEGRAL data, a re-analysis
of archival BeppoSAX data, and the spectral analysis of XMM-Newton data.


\section{Observations}
\subsection{BeppoSAX data}
\citet{sax}, hereafter {\it Au03},  reported the discovery of the serendipitous source SAX
J1802.7-2017, identified as an X-ray pulsar.  This result was from
BeppoSAX archival observations of the GX 9+1 field performed from
September 16 2001 (02:01:30.0 UTC) to September 20 2001 (03:00:08.5 UTC).
The position of SAX J1802.7-2017 was given as R.A. (2000.0) =
18$^{h}$02$^{m}$39.9$^{s}$ and Dec. = -20$^{\circ}$17$\arcmin$13.5$\arcsec$ with a
positional uncertainty of 2$\arcmin$.  A pulse
period of 139.612 s and an orbital period of $\sim$4.6 days was
reported. Table~\ref{tab:sax_fit} shows all of their measured system
parameters.  The total duration of the observation $\sim$4.0 days, was
such that the data did not cover a complete orbit.  As a result the
measurement of the orbital period through the
delays in pulse arrival times was assocated with a 9\% error which
contributed to other system parameters having errors of $\sim$30\%.
The possibility that this was an eclipsing system was also proposed.

   \begin{table}[tbhp]
   \begin{center}
     \caption{Estimated phase delay fit parameters and system
     parameters of SAX J1802.7-2017 ({\it Au03}).  [$a_{0}$, $a_{1}$,
     and $B$ are phase delay fit parameters defined in
     Section 3.2.]}
    \begin{tabular}[h]{ll} \hline
      Parameter & Value \\
      \hline \hline
      $a_{0}$  &  0.04$^{+0.12}_{-0.09}$ \\
      $a_{1}$  &  0.02 $\pm$ 0.03 days$^{-1}$ \\
      $B$        &  0.50$^{+0.07}_{-0.05}$ \\
      $P_{orb}$ & 4.6$^{+0.4}_{-0.3}$ days \\
      $a_{x}\sin{i}$        &  70$^{+10}_{-7}$ lt-s \\
      $T_{\pi/2}$ & 52168.22$^{+0.10}_{-0.12}$ MJD \\
      $P_{pulse}$ & 139.612$^{+0.006}_{-0.007}$ s \\
      $f(M)$ & 17 $\pm$ 5 $M_{\sun}$ \\
      $\theta_{\epsilon}$   & 0.64 $\pm$ 0.14 rad \\   
      $M_{C}$  &  $\gtrsim$ 11 $M_{\sun}$  \\
      $R_{C}$  &   $\gtrsim$ 14 $R_{\sun}$  \\
      \hline
      \end{tabular}
     \label{tab:sax_fit}
     \end{center}
   \end{table}

\subsection{INTEGRAL data}

IGR J18027-2016 has been observed on multiple occasions during the
core programme by the IBIS detector ISGRI [INTEGRAL Soft Gamma-Ray
Imager, \citep{isgri}].  IBIS generates images of the sky with
$\sim$30$^{\circ}$ field of view in the energy range 15-1000 keV \citep{ibis_fov}.
IBIS observed
IGR J18027-2016 over three epochs: 28 Feb. -- 28
Apr. 2003; 27 Sep. -- 12 Oct. 2003; 17 Feb. -- 20 Apr. 2004.  ISGRI
images were generated for each
selected pointing in 10 narrow energy bands with the ISDC Offline
Scientific Analysis (OSA) software version 4.1.  The individual images were
combined to produce mosaics of the region in broader energy bands
using the system described in \citet{ibissurvey}. 

\begin{figure}[htbp]
	\centering
	\includegraphics[width=0.9\linewidth, clip]{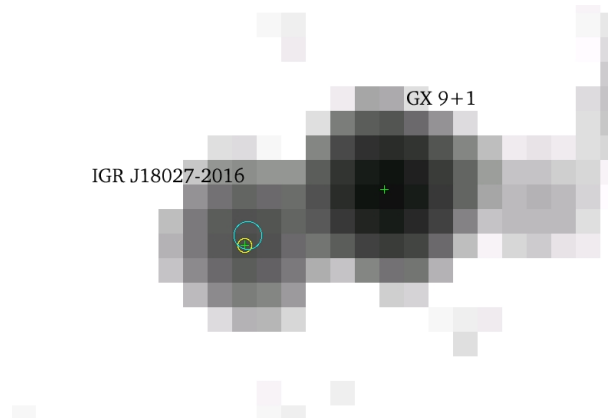}
	\caption{ISGRI 20-40 keV image of the field containing IGR
	J18027-2016 and GX 9+1.  The crosses correspond to the source
	positions; the larger circle is the BeppoSAX error circle \citep{sax}; the
	smaller circle is the ISGRI 90$\%$ error circle \citep{psle}.}
		\label{image}
	\end{figure}

Fig.~\ref{image} shows the IBIS/ISGRI 20-40 keV image of the region
around IGR J18027-2016.  The atoll source, GX 9+1
is also clearly visible $\sim$22$\arcmin$ from IGR J18027-2016.  IGR
J18027-2016 appears as a 30.5$\sigma$ detection in the 20-40 keV band
at R.A. (2000.0) = 18$^{h}$02$^{m}$46.1$^{s}$ and Dec. =
-20$^{\circ}$17$\arcmin$37.1$\arcsec$ with a positional uncertainty of $\sim$1$\arcmin$
\citep{psle}.  This position is consistent with the reported position of SAX
J1802.7-2017 ({\it Au03}).  The source flux in the 20-40 keV energy
band is F$_{20-40keV}$ = 0.58 $\pm$ 0.02 counts sec$^{-1}$ which
corresponds to $\sim$6.4 mCrab.

\subsection{XMM data}
IGR J18027-2016 was observed by XMM-Newton for $\sim$12 ksec,
from 06:54:40.0 -- 09:44:38.0 UTC, 2004 April 06.  A single X-ray
source was found within the ISGRI error circle in the EPIC PN and MOS
cameras \citep{xmm-epic, xmm-mos}.  The XMM Science Analysis System
(SAS), software version 6.0, was used to extract light curves and
spectra for IGR J18027-2016. Standard SAS tools were used to calculate
the instrumental response and the effective area for the extracted spectra.

\section{Temporal Analysis}
\subsection{ISGRI}

	\begin{figure*}[btp]
	\centering
	\includegraphics[width=0.9\linewidth, clip]{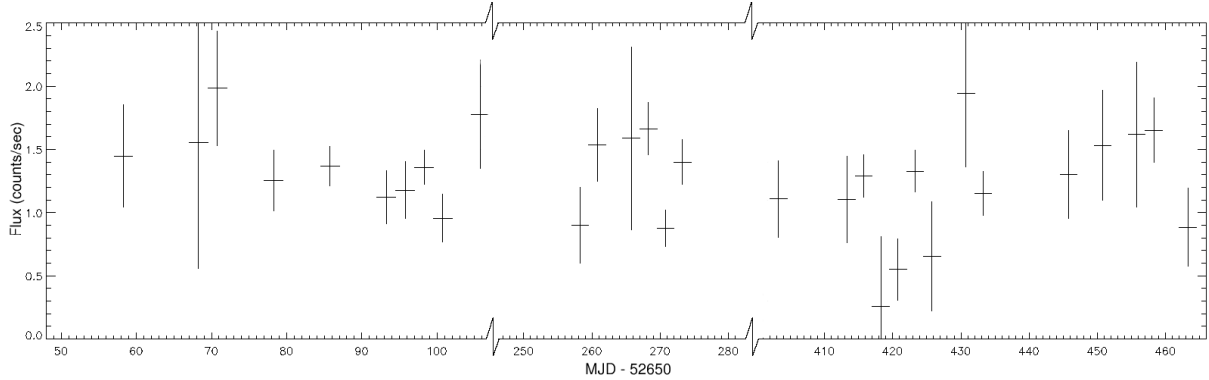}
	\caption{20-80 keV light curve of IGR J18027-2016 covering the
	three observed epochs. The light curve has been rebinned to
	2.5 day averages.}
		\label{lightcurve}
	\end{figure*}

\begin{figure}[tbp]
   \centering
   \includegraphics[width=1.0\linewidth, clip]{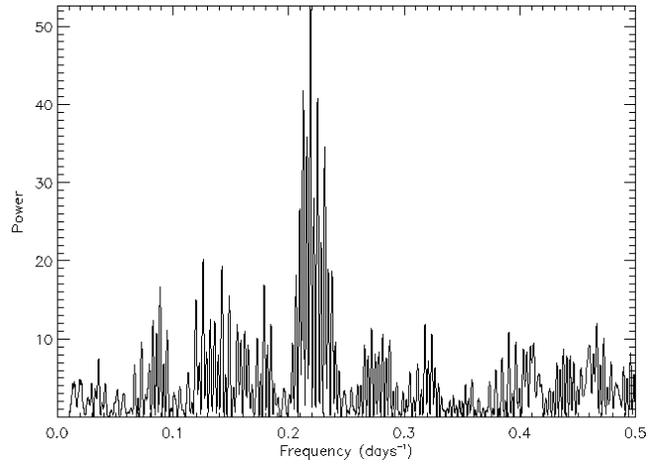}
   \caption{Lomb-Scargle periodogram generated from the 20-40 keV
   light curve of IGR J18027-2016.}
              \label{powspec}%
    \end{figure}

The flux of IGR J18027-2016 was extracted
from each ISGRI pointing image to generate light curves in different
energy bands, the rebinned 20--80 keV light curve spanning $\sim$417 days is
seen in Fig.~\ref{lightcurve}.  The source is not especially strong
above 40--50 keV and to minimize systematic effects we searched the
20--40 keV light curve for periodicities using the
Lomb-Scargle periodogram method by means of the fast implementation of
\citet{fast_lsp}.  The resulting power spectrum is
shown in Fig.~\ref{powspec}.  The peak power of 52.58 corresponds to a
frequency of 0.2188 days$^{-1}$.  The error on the angular frequency
\citep{period_err} is:

\begin{equation}
	\delta\omega = \frac{3 \pi \sigma_{N}}{2 \sqrt N T A}
\end{equation}

where $\sigma$$_{N}$$^{2}$ is the variance of the noise, N is the
number of data points, T is the
total length of the data set and A is the amplitude of the signal given by:

\begin{equation}
	A = 2 \sqrt{\frac{z_{0} \sigma_{s}^{2}}{N}}
\end{equation}
where z$_{0}$ is the Lomb-Scargle power and $\sigma$$_{s}$$^{2}$ is the variance of the
light curve.

This frequency corresponds to a period of 4.570 $\pm$ 0.003 days which is
consistent with the orbital period indirectly measured by {\it Au03}.

The significance of this peak was confirmed by applying a randomization
test.  The time stamps for each flux measurement were
randomly reordered and the periodogram of the resulting light curve
generated.  We simulated 5000 light curves in this fashion.  The mean
maximum power found in the randomized data sets was 10.6 and highest
power recorded was a single instance of 19.5.  It is immediatley
obvious that a peak of power $\sim$53 is extremely unlikely to occur
by chance and hence our detection is highly significant.

   \begin{figure}[tbp]
   \centering
   \includegraphics[width=1.0\linewidth, clip]{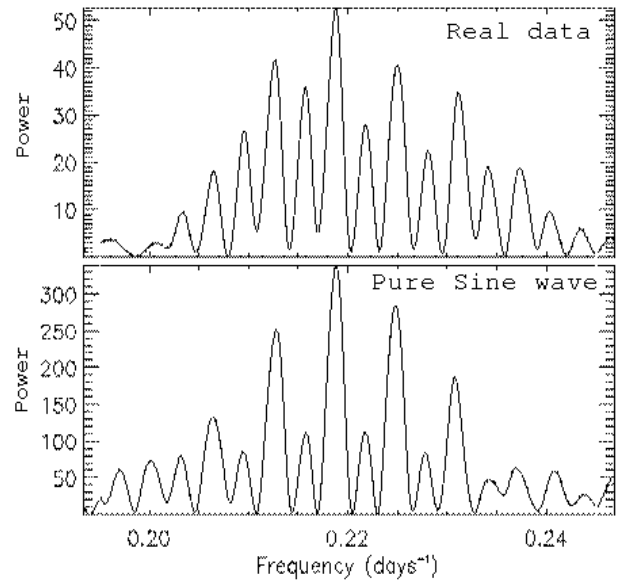}
   \caption{{\it Top}: Zoom on the peak of the periodogram of IGR
   J18027-2016, shown in Fig.~\ref{powspec}.  {\it Bottom}:
   Lomb-Scargle periodogram generated from a sine wave with
   a period of 4.570 days and sampled identically to IGR J18027-2016.}
              \label{fake}%
    \end{figure}

Upon close examination of the periodogram it can be seen that there
is some aliasing around the frequency of the detected period.
This raises the question of whether we have selected the correct peak
and hence the correct frequency/period.  The much higher power of our
selected frequency is sufficient that we can be confident that this is
the true frequency of the system.  To investigate the orgin of the
aliasing pattern a pure sine wave of our
chosen frequency was sampled in an identical fashion to that of the
original data set.  The corresponding periodogram (Fig.~\ref{fake})
can be seen to exhibit broadly the same
aliasing pattern as observed in the data.  The peak of maximum power
corresponds to the frequency of the input sine wave and the aliases
are spaced in an identical fashion to that of the original data set.
Also, as discussed in section 3.2, folding the ISGRI light curve on a
period of 4.570 days indicates an eclipse of the system exactly when
we would expect it to be observed.  Hence, we are confident
that the peak of maximum power in our data corresponds to the true
frequency of the system.  

\subsection{BeppoSAX}
We used the orbital period as measured by ISGRI to re-analyse the
BeppoSAX data presented by {\it Au03}.  Originally, phase delays in
the pulse period of the pulsar measured by SAX were fitted without any
prior knowledge of the orbital period.  We performed an identical
analysis of the phase delays assuming an orbital period of 4.570 days.

\begin{figure}[bhtp]
   \centering
   \includegraphics[width=0.9\linewidth, clip]{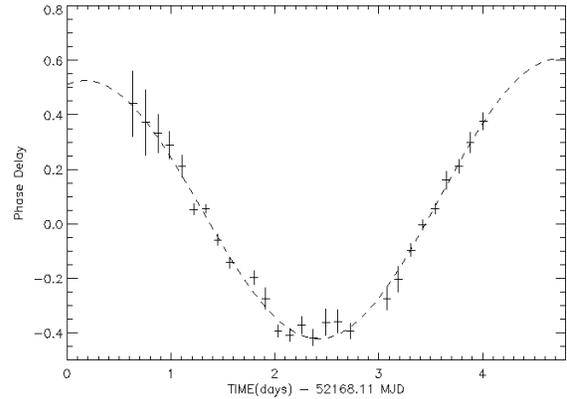}
   \caption{Phase delays as a function of time as measured by
   BeppoSAX.  The dashed line is the best-fit function (Eq.[\ref{phase_delay}]).}
              \label{sax_phase}%
    \end{figure}

The modulation of the phase delays can be explained by the propagation
delays resulting from the orbital motion of the neutron star around
its companion star.  The phases were fit with

\begin{equation}
  \Delta\phi = a_{0} + a_{1}t_{n} + B\cos{\left[\frac{2\pi(t_{n} - T_{\pi/2})}{P_{orb}}\right]}
\label{phase_delay}
\end{equation}

where $t_{n}$ is the arrival time of the {\it n}th pulse, $T_{\pi/2}$
is the epoch of NS superior conjunction of the NS, the linear
term $a_{1}$ = $\Delta$$P_{pulse}$$/$$P_{pulse}^2$ and $P_{orb}$ =
4.570 $\pm$ 0.003 days.  The $\chi$$_{\nu}^2$ of the fit was 1.3, the
fit parameters are listed in Table~\ref{tab:phase_fit} and the fit is
shown in Fig.~\ref{sax_phase}.  As $a_{x}$$\sin${\it i} =
$P_{pulse}B$, then $a_{x}$$\sin${\it i} $\sim$ 68 $\pm$ 1
lt-s; $P_{pulse}$ $\sim$139.612 s ({\it Au03}).  All of the fit
parameters are consistent with those originally measured by
{\it Au03}, however they are considerably better constrained due to the
independent and precise measurement of $P_{orb}$ by IBIS/ISGRI. 

   \begin{table}[htbp]
   \begin{center}
     \caption{Phase delay fit parameters}
    \begin{tabular}[h]{ll} \hline
      Parameter & Value \\
      \hline \hline
      $a_{0}$  &  0.03 $\pm$ 0.07 \\
      $a_{1}$  &  0.02 $\pm$ 0.03 days$^{-1}$ \\
      $B$        &  0.49 $\pm$ 0.003 \\
      $T_{\pi/2}$ & 52168.26 $\pm$ 0.04 MJD \\
      \hline
      \end{tabular}
     \label{tab:phase_fit}
     \end{center}
   \end{table}

\begin{figure}[tbp]
   \centering
   \includegraphics[width=1.0\linewidth, clip]{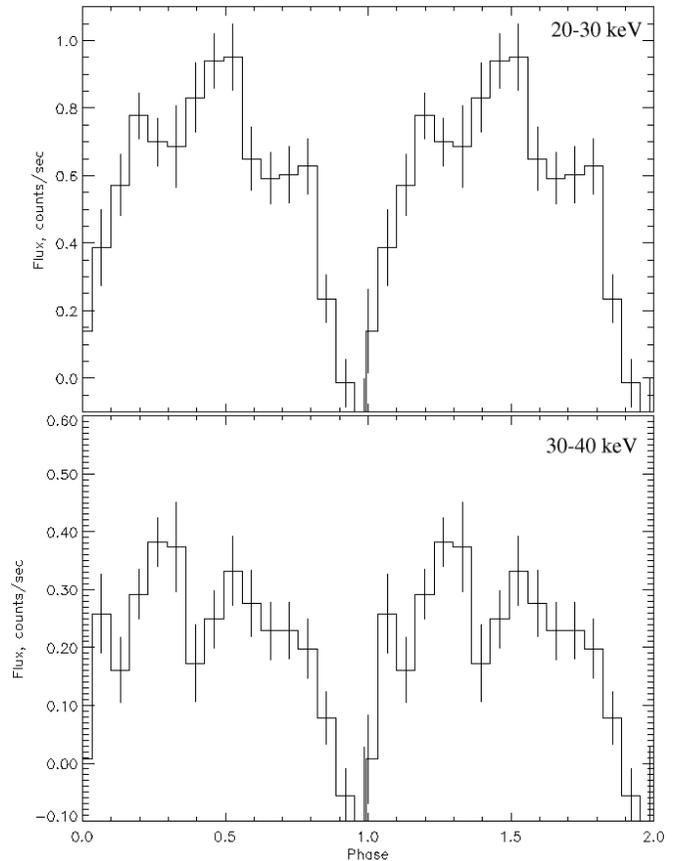}
   \caption{Folded light curves in the 20-30 and 30-40 keV energy bands of IGR
   J18027-2016. The
   folding is performed for an orbital period of 4.570 days, and the
   zero epoch was assumed at the superior conjunction of the
   NS,$T_{\pi/2}$ = 52168.26 MJD .}
              \label{lc}%
    \end{figure}

Using the measured orbital period and the epoch of superior
conjunction of the NS, $T_{\pi/2}$, obtained in the fit, the IBIS/ISGRI light
curves were phase-folded in the 20-30 keV and 30-40 keV energy bands.
These are shown in Fig.~\ref{lc} and clearly show a sharp eclipse as
suggested by {\it Au03}.  However, the eclipse evident in
Fig.~\ref{lc} appears assymmetic and indicates that we may be
partially resolving the eclipse suggesting that the hard X-ray
emission may not exclusively originate from a point source.  The
fuller phase coverage of the IBIS/ISGRI data unambiguously confirms
the orbital period and the presence of an eclipse in the system.  The
time of mid-eclipse occurs at the time of superior conjunction of the
NS, $T_{\pi/2}$.

\subsection{Refining the orbital period \& an accurate ephemeris}
Taking the zero epoch as the time of superior conjunction of the NS, $T_{\pi/2}$
= 52168.26 $\pm$ 0.04 MJD and the orbital period, $P_{orb}$ =
4.570 $\pm$ 0.003 days the number of orbital cycles which have
occured by the time of the last ISGRI observation is E =
207.2 $\pm$ 0.1.  Hence, over this timescale, we can confidently
identify which cycle we observe.  This is borne out by Fig.~\ref{lc},
as the eclipse is evident very near zero phase when the ISGRI data is
folded with the BeppoSAX zero epoch.  From the folded ISGRI light
curves the time of an eclipse around the mid point of the ISGRI
observation is calculated, $T_{eclipse}$ = 52931.4 $\pm$ 0.2 MJD.
This would correspond to the beginning of the 167$^{th}$ orbital
cycle.  Linearly fitting between these two times of eclipse
measurements yields a refined orbital period estimate of  $P_{orb}$ = 4.5696 $\pm$
0.0009 days.  This allows the calculation of an accurate mid-eclipse
ephemeris from:
\begin{equation}
  T_{mid} = T_{0} + P_{orb} \times E
\label{ephemeris}
\end{equation}
where T$_{0}$ is the time of the 1$^{st}$ eclipse seen by BeppoSAX and
E is the cycle number.  This gives an ephemeris of, $T_{mid}$ =
52931.37 $\pm$ 0.04 MJD.

\subsection{XMM}

\begin{figure}[b]
\centering
\includegraphics[width=1.0\linewidth, clip]{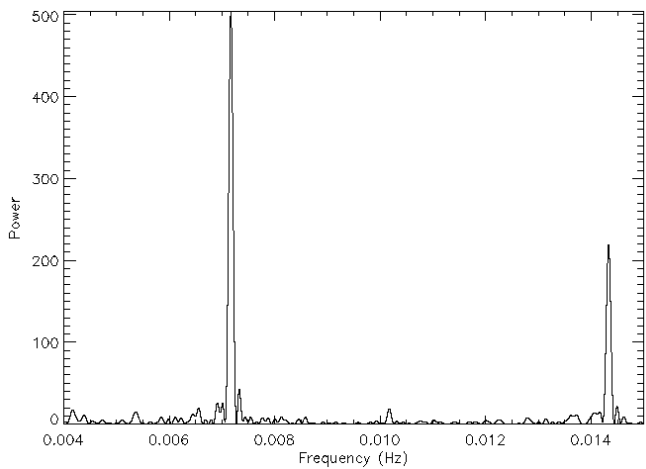}
   \caption{The Lomb-Scargle periodogram generated from the XMM EPIC
   2--10 keV light curve of IGR J18027-2016.}
              \label{xmm_ps}%
    \end{figure}

The timing mode of the PN camera allows the collection of timing data
with a resolution of 30 $\mu$s \citep{xmm-epic}.  The arrival time of
all events were corrected to the solar system barycentre, using the
SAS task ``barycen''.  A 2--10 keV band light curve consisting of 1
second bins was generated for the duration of the observation.
The Lomb-Scargle periodogram method performed on the ISGRI data was
applied to search for periodicites.  The resulting power spectrum is
shown in Fig.~\ref{xmm_ps}.  A peak of power $\sim$504 is
clear at a frequency of 7.17 mHz, the second harmonic of this frequency is
also clearly visible with a power of $\sim$220.  This corresponds to
a pulse period of, $P_{pulse} =$ 139.47 $\pm$ 0.04 seconds.  This is
consistent with the range of pulse periods (139.44 -- 139.86 s) detected
by {\it Au03}.  At the time of the XMM observation we expect the
system to be at a phase of $\sim$0.2 in its orbital cycle.  Hence, the
pulse period should be doppler shifted by the orbital motion.
Correcting the measured pulse period to the rest frame of the pulsar
yields, $P_{pulse} =$ 139.61 $\pm$ 0.04 seconds.  The XMM measurement
is consistent with the pulse period, $P_{pulse} =$ 139.612 $\pm$ 0.006
seconds, measured by BeppoSAX.

   \begin{figure}[t]
   \centering
   \includegraphics[width=0.9\linewidth, clip]{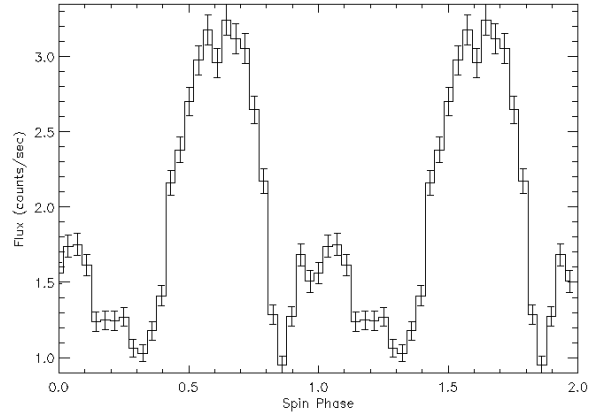}
   \caption{Pulse period phase folded light curve in the 2--10 keV
   energy band.  The folding is obtained for a pulse period of 139.47
   seconds and the zero epcoh is taken at the superior conjunction of the NS,
   $T_{\pi/2}$ = 52168.26 MJD.}
              \label{xmm_pulse}%
   \end{figure}

Folding the XMM data on this pulse period yields the pulse
period phase folded light curve seen in Fig.~\ref{xmm_pulse}.  The
main pulse occurs around phase 0.63.  A secondary peak is visible
around phase 0.05.  The pulse fraction, $(I_{max} - I_{min})/I_{max}$,
where $I_{max}$ and $I_{min}$ are the maximum and minimum count rates,
is $\sim$68\% $\pm$ 10\%.  The pulse period phase folded light curve
matches well with the folded light curves of {\it Au03}, the pulse
fractions are consistent and the locations of the primary and
secondary peaks also agree.

   \begin{figure*}[hbtp]
   \centering
   \includegraphics[width=0.9\linewidth, clip]{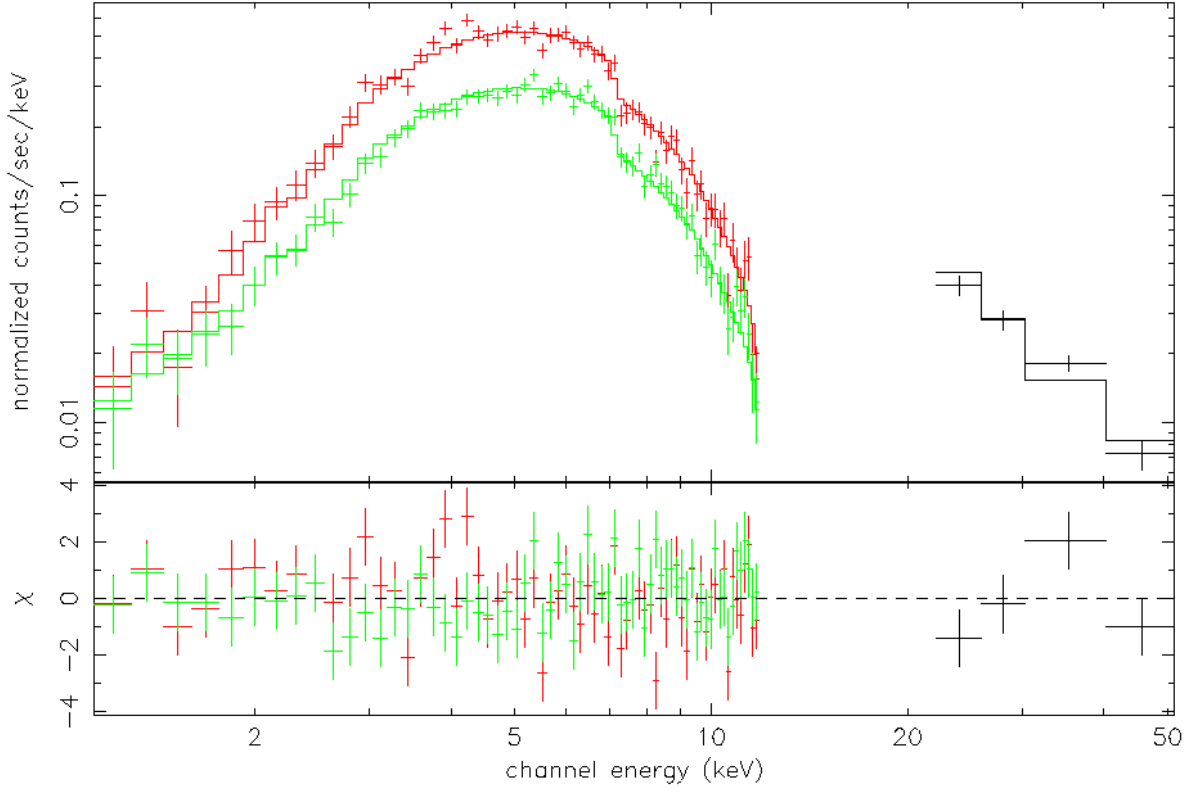}
   \caption{The phase resolved XMM ({\it upper left: EPIC PN
   spectrum of primary pulse.  Lower left: EPIC PN spectrum outside of
   the primary pulse}) and INTEGRAL photon spectrum ({\it right: ISGRI
   spectrum})of IGR J18027-2016
   with the best fit model.}
              \label{spectrum}%
   \end{figure*}

\section{Spectral Analysis}
The ISGRI spectrum was extracted from narrow energy band
mosaics and hence represent the long term average spectrum of the
source.  Phase resolved spectra were extracted from the XMM-EPIC data set.
Using the pulse profile (Fig.~\ref{xmm_pulse}) a spectrum of the
source during the primary pulse and a spectrum outside of the primary
pulse were generated.  {\it Au03} reported that the low statistics of
the BeppoSAX data did not allow for an accurate spectral analysis.
Additionally, the BeppoSAX PDS data is difficult to analyse as the
data is contaminated by the bright atoll source, GX 9+1 which also
appears in the field of view.

The EPIC PN and ISGRI average spectra were simultaneously fit, in XSPEC
v11.3, by a photoelectric absorbed, broken power law model with a single Gaussian
emission line (gauss + zvphabs*bknpower).  A broad Gaussian line is used
to account for soft X-ray residuals that mostly concentrate at low
energy.  This soft excess contributes 7 $\times 10^{-14}$ ergs
cm$^{-2}$ s$^{-1}$ below 3 keV.  Elemental abundances of the spectral model, excluding Fe, are frozen to
solar quantities and the redshift is frozen to 0, all other parameters
are free.  The fit has $\chi$$_{\nu}^2$ = 1.29.  The best fit parameters are listed
in Table~\ref{tab:xmm_fit}.  The broken power law indices were found to
be $\Gamma_{1}$ = 0.79 $\pm$ 0.08 and $\Gamma_{2}$ = 3.1 $\pm$ 0.1,
with a break at $E_{Break}$ = 11.3 $\pm$ 0.6 keV.  A column density of
$N_{H}$ = 6.8 $\pm$ 1.0 $\times 10^{22}$ cm$^{-2}$ is found compared to
the expected galactic column density of 1.0 $\times 10^{22}$ cm$^{-2}$.  The Fe
edge is $\sim$3 times that of solar abundances, this may indicate that
large amounts of Fe are present or that the $N_{H}$ is
underestimated.  The shape of the spectrum is typical of
X-ray pulsars which are characterized by a flat power law with a slope
of between 0 and 1 up to a high energy cut-off at 10-20 keV after which the slope
becomes much steeper \citep{pulsar_spec}.  Fitting Fe lines at 6.45
and 7.1 keV improves the overall fit, however the F test indicates
that they are significant at only the 95\% confidence level.  The ISGRI
spectrum flux is, F$_{20-100 keV}$ = 5.9 $\times 10^{-11}$ ergs cm$^{-2}$ s$^{-1}$, the
XMM unabsorbed fluxes are, F$_{2-10 keV}$ = 8.9 $\times 10^{-11}$ ergs
cm$^{-2}$ s$^{-1}$ within the primary pulse, and  F$_{2-10 keV}$ = 5.1 $\times 10^{-11}$ ergs
cm$^{-2}$ s$^{-1}$ outside the primary pulse.

   \begin{table}[tbhp]
   \begin{center}
     \caption{Best fit spectral parameters (90\% confidence) of the
     spectral model discussed in the text and simultaneously fit to
     the XMM-EPIC and INTEGRAL-ISGRI spectra.}
    \begin{tabular}[h]{ll} \hline
      Parameter & Value \\
      \hline \hline
      $\chi^{2}/\nu$  &  164.57/128 \\
      $N_{H}$  &  6.8 $\pm$ 1.0 $\times$ 10$^{22}$ cm$^{-2}$ \\
      $\Gamma_{1}$        &  0.79 $\pm$ 0.08  \\
      $E_{Break}$        &  11.3 $\pm$ 0.6 keV  \\
      $\Gamma_{2}$        &  3.1 $\pm$ 0.1  \\
      Bkn power law $I_{1 keV}$ & 4.9 $\pm$ 1.0 $\times$ 10$^{-3}$
      ph keV$^{-1}$ cm$^{-2}$ s$^{-1}$ \\
      Fe abundance & 3.1 $\pm$ 0.8 (of solar) \\
      \hline
      \end{tabular}
     \label{tab:xmm_fit}
     \end{center}
   \end{table}

The shapes of the phase resolved XMM spectra are consistent with each other.  The
hydrogen column density and continuum parameters of the model,
excluding the normalization, do not vary with pulse phase indicating
that the absorption is not associated with the neutron star accretion column.

It is possible to fit the data with a Comptonization model or a
cut-off power law, however to do so requires the introduction of a particularly small 
normalization constant to the spectral model, C$_{ISGRI}$$\sim$0.2, to
account for cross-calibration uncertainties of the ISGRI and EPIC
observations.  This would indicate that XMM had observed IGR J18027-2016 while it was in a high
state when compared to the average ISGRI flux.  However, the ISGRI
light curve (Fig.~\ref{lightcurve}) at the time of the XMM observation
show no indication of a large change in flux of the source.  The
fitting of system $N_{H}$ is highly dependent upon the spectral shape
of the continnum assumed in the spectral model, a
softer spectrum (e.g. the comptt model describing Comptonization of
soft photons in a hot plasma) results in an $N_{H}$ twice as large as
listed in Table~\ref{tab:xmm_fit}.

   \begin{figure*}[thbp]
   \centering
   \includegraphics[width=0.9\linewidth, clip]{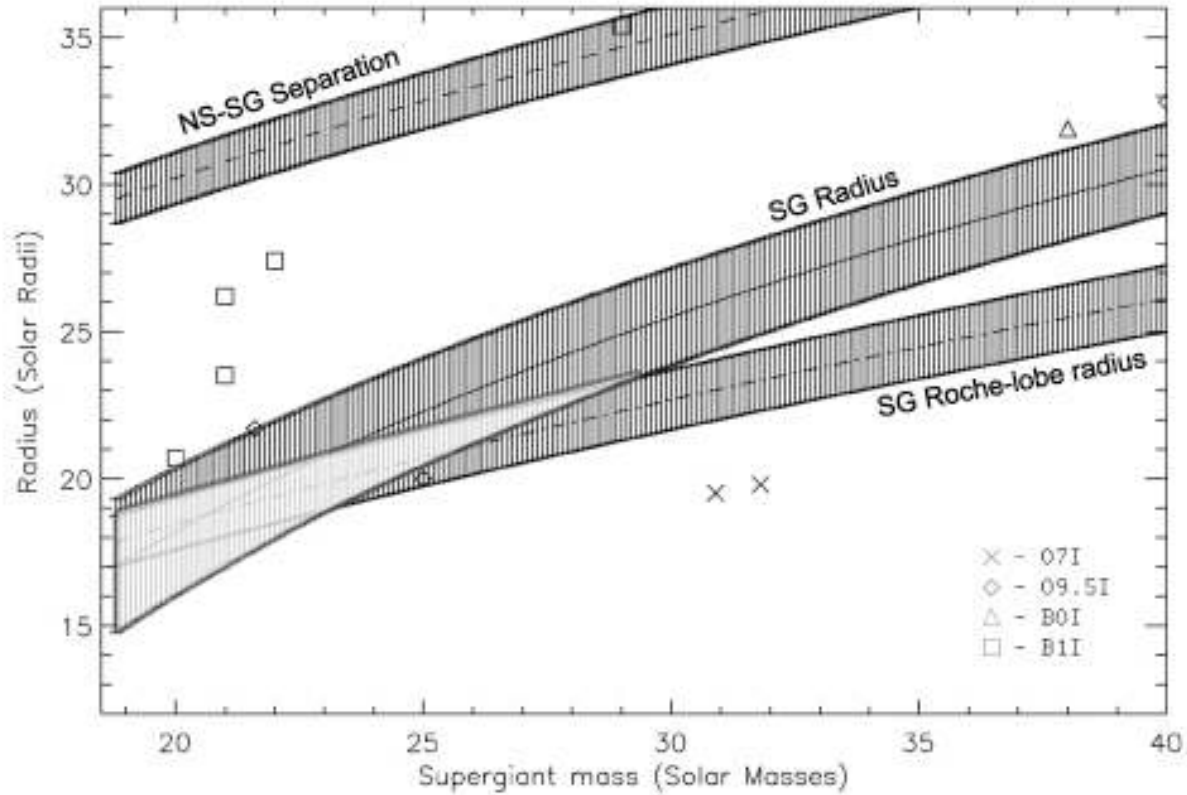}
   \caption{The derived mass-radius relationship of the supergiant (SG) in
   IGR J18027-2016 (assuming a NS mass of 1.4 $M_{\sun}$) is plotted
   against the SG roche lobe radius and the NS-SG separation.  The
   lighter shaded area corresponds to the region in which the SG
   radius is smaller than its roche lobe radius.  Also plotted are
   the masses and radii of a number of known OB supergiants.}
              \label{mass_radius}%
    \end{figure*}

\section{Discussion}
We have directly and precisely  measured the orbital period for IGR
J18027-2016 of 4.5696 $\pm$ 0.0009 days.  This is consistent
with the implied orbital period of SAX J1802.7-2017, and as it is the
only known X-ray source within 10$\arcmin$ according to the SIMBAD/NED
database the two sources are confirmed as the same
source.  Additionally, we have observed that this is an eclipsing
system confirming the hypothesis of {\it Au03}.  The ephemeris of the
mid-eclipse time is measured as, $T_{mid}$ = 52931.37 $\pm$ 0.04 MJD.
The $\sim$139 s pulse period measured by BeppoSAX and XMM combined
with the 4.5696 day orbital period, situates IGR J18027-2016 in the
region of the Corbet diagram \citep{Corbet} occupied by underfilling
Roche lobe supergiant systems .  Hence, the system appears to be a
neutron star that is accreting material through the stellar
wind of the companion supergiant star.  

The ISGRI 20-80 keV light curve,
shown in Fig.~\ref{lightcurve}, indicates that this is a weak,
persistent source with no evidence of flaring in any
of our observations.  \citet{luto_gc} also reported that the 18--60
keV fluxes observed by IBIS in September 2003 and March 2004 of IGR
J18027-2016 were
nearly constant and equal.  The absence of strong hard X-ray variability
contrasts with what has been observed in other HMXBs in which flares
by a factor of 10 have been observed.  This may indicate that the
stellar wind is rather homegenous at least on time scales longer than
a few days.  The persistent nature of the source suggests that the NS is continuously
accreting, and, combined with the sinusoidal modulation of the pulse
arrival times this implies that the NS is in an approximately circular
orbit about its companion.  This is consistent with the
eccentricity upper limit of e~$\lesssim$~0.2 given by {\it Au03}.  The
absence of apparent transient behaviour and circular orbit further
suggests that this is a supergiant system and not a Be X-ray binary
\citep{HMXB_Liu}.  However, the fact that IGR J18027-2016 has not been
reported by previous X-ray telescopes may indicate some level of transient
behaviour.  Notably there was no detection by the ROSAT mission.  Transient
behaviour would indicate the absence of Roche lobe overflow but would
suggest that this was an atypical Be X-ray binary as its orbital and
pulse period do not locate it in the expected region of the Corbet
diagram and it has a very low eccentricity orbit.

During the XMM observation the X-ray flux increased
by 50\% with constant pulse fraction.  Spectra derived for both
periods do not show any significant variation in the spectral
parameters.  This is notable as one could expect to see a significant
increase in $N_{H}$ for large flux differences.  The observed
properties of IGR J18027-2016 appear to be very stable on timeescales
of a few days, as seen by ISGRI, and 12 ksec, as seen by XMM.

The mass function of the system donor is:

\begin{equation}
  f(M) = \frac{M_{C}\sin^2{i}}{(1 + q)^2} =
  \left(\frac{2\pi}{P_{orb}}\right)^2 \frac{(a_{x}\sin{i})^3}{G}
  \sim16 \pm 1 M_{\sun}
\label{mass_func}
\end{equation}

where $M_{C}$ is the supergiant companion star mass, $q = \frac{M_{x}}{M_{C}}$, is the ratio of
the NS mass to the supergiant companion mass and $G$ is the
gravitational constant.

Fig.~\ref{lc} clearly shows an eclipse occuring in the 20--30 and
30--40 keV energy bands.  Future observations of the ingress and
egress of the NS eclipse may provide information regarding the donor
star's atmosphere and indicate any evidence of extended high energy
emission in the region around the NS.  There are two methods to determine the half
angle of eclipse, $\theta_{\epsilon}$, which we define as the half
width half height of the eclipse.  Firstly, if we assume a
cicular orbit, the measured
duration of the eclipse corresponds to a half angle of eclipse of
$\theta_{\epsilon}$ = 0.66 $\pm$ 0.2 radians in both energy bands.
Additionally, the mid-eclipse time corresponds to the epoch of NS
superior conjunction of the NS, $T_{\pi/2}$, derived from our fit.  This
corresponds to the period in the BeppoSAX data when the count rate was
consistent with the background count rate.  Taking the time at which
the eclipse ends as 52,168.69 MJD ({\it Au03}), then the half-angle of
the eclipse corresponds to $\theta_{\epsilon}$ = 0.591 $\pm$ 0.09
radians.  Taking a weighted mean of the half angle of eclipse gives
$\overline{\theta}_{\epsilon}$ = 0.61 $\pm$ 0.08 radians.  Assuming we
have a point source and spherical geometry the
duration of the eclipse is related to $R_{C}$, the companion star
radius, the inclination of the system and $a$, the separation of the
stars by \citep{nagase}:

\begin{equation}
  \frac{R_{C}}{a} = (\cos^2{i} +
  \sin^2{i}\sin^2{\overline{\theta}_{\epsilon}})^{1/2}
\label{radius_func}
\end{equation}
               
Assuming that the NS has the canonical mass of 1.4 M$_{\sun}$ then for any
mass of the companion star the inclination of the orbit can be dervied
from the mass function (eq.~\ref{mass_func}).  Assuming that we
have a perfectly circular orbit then $a_{x}$$\simeq$$a$, the
phase delay modulation yields
$a_{x}\sin{i}$ and hence from eq.~\ref{radius_func} there is an
empirical relationship between the companion star mass and radius.
An analytic approximation for the companion star Roche lobe is:

\begin{equation}
  R_{Roche} = a\left(0.38 + 0.2\log{\frac{M_{C}}{M_{NS}}}\right)
\label{roche_eq}
\end{equation}

\citep{roche_rad}   (accurate $<$ 1$\%$)

From its low luminosity and its position in the Corbet diagram, we
expect the supergiant
companion star to be underfilling its Roche lobe hence the Roche lobe
radius gives an upper limit to the allowed companion star radius, mass
and system inclination.  The empirical $M_{C}$--$R_{C}$ relation and the
associated $R_{Roche}$ are plotted in Fig.~\ref{mass_radius} along
with the neutron star -- supergiant separation.  The lighter shaded
region towards the lower left of Fig.~\ref{mass_radius} indicates the
allowed system parameters such that our measurements represent an
underfilling Roche lobe supergiant.  Also plotted are the masses and
radii of a number of known OB supergiants \citep{OB1, OB2}.

From Fig.~\ref{mass_radius} we can see that the properties of the
companion star are: $M_{C}$ $\sim$ 21$M_{\sun}$; $R_{C}$ $\sim$
19$R_{\sun}$.  The complete range of these values is given in
Table~\ref{tab:sys_par} with the other parameters of the system.
These mass and radius values are consistent with late O9 -- early B1
supergiant stars. 

   \begin{table}[t]
   \begin{center}
     \caption{System parameters of IGR J18027-2016}
    \begin{tabular}[h]{ll} \hline
      Parameter & Value \\
      \hline \hline
      $P_{orb}$  &  4.5696 $\pm$ 0.0009 days \\
      $P_{pulse}$ (BeppoSAX)  &  139.612 $\pm$ 0.006 s \\
      $P_{pulse}$ (XMM)  &  139.61 $\pm$ 0.04 s \\
      $a_{x}\sin{i}$        &  68 $\pm$ 1 lt-s \\
      $T_{\pi/2}$ & 52168.26 $\pm$ 0.04 MJD \\
      $T_{mid}$ & 52931.37 $\pm$ 0.04 MJD \\      
      $f(M)$ & 16 $\pm$ 1 $M_{\sun}$ \\
      $\overline{\theta}_{\epsilon}$   & 0.61 $\pm$ 0.08 rad \\   
      $M_{C}$  &  18.8 $\lesssim$ $M$ $\lesssim$ 29.3
      $M_{\sun}$  \\
      $R_{C}$  &  14.7 $\lesssim$ $R$ $\lesssim$ 23.4
      $R_{\sun}$  \\
      $a$  &  28.4  $\lesssim$ $a$ $\lesssim$ 35.8
      $R_{\sun}$  \\
      $i$  &  59$^\circ$ $\lesssim$ $i$ $\lesssim$ 88$^\circ$  \\
      \hline
      \end{tabular}
     \label{tab:sys_par}
     \end{center}
   \end{table}

   \begin{table}[tbhp]
   \begin{center}
     \caption{Theoretical $\dot{P}_{pulse}$ and distance for
     different source luminosity}
    \begin{tabular}[h]{|l|llll|} \hline
      Luminosity, (ergs cm$^{-2}$ s$^{-1}$)  &  10$^{35}$ & 10$^{36}$ & 10$^{37}$ & 10$^{38}$\\
      \hline
      $\dot{P}_{pulse}$, (s yr$^{-1}$)  &  0.02 & 0.2 & 1.4 & 9.7 \\
      Distance, (kpc)        &  3.1 & 9.7 & 31 & 421\\
      \hline
      \end{tabular}
     \label{tab:spin_l}
     \end{center}
   \end{table}

Measurement of changes in the NS pulse period is an independent
indicator of the accretion mechanism operating in the system.  If the
NS is accreting matter from the donor star via Roche lobe overflow the
NS should show evidence of spinning up.  The BeppoSAX and XMM
measurements of the spin period as 139.612 $\pm$ 0.006 s and 139.61
$\pm$ 0.04 s respectively, are separated by $\sim$2.5 years and
indicate no evidence of a change in spin period.  Taking the 1$\sigma$
limits of these measurements we find an upper limit to the
spin-up rate of, $\dot{P}_{pulse}$$\sim$0.02 s yr$^{-1}$.  The
theoretical spin-up rate is given by \citep{spin_up}:

\begin{equation}
  \frac{-\dot{P}_{pulse}}{P_{pulse}} \cong 8 \times 10^{-5}
  M_{1}^{-3/7} R_{6}^{6/7} L_{37}^{6/7} \mu_{30}^{2/7}
  I_{45}^{-1}P_{pulse}  yr^{-1}
\label{spin}
\end{equation}

where M$_{1}$ is the NS mass in M$\sun$, R$_{6}$ is the NS radius in
cm, L$_{37}$ is the NS accretion luminosity in units 10$^{37}$ ergs cm$^{-2}$
s$^{-1}$, $\mu_{30}$ is the magnetic moment in units 10$^{30}$ G
cm$^{3}$ and I$_{45}$ is the moment of inertia ($\sim$MR$^{2}$) of the
NS in units 10$^{45}$ g cm$^{2}$.  For a typical neutron star with a
magnetic field strength, B$\sim$10$^{12}$ G, $\mu_{30}$ $\sim$
I$_{45}$ $\sim$ R$_{6}$ $\sim$ 1.  The theoretical spin-up of a
neutron star accreting via Roche lobe overflow with a luminosity in
the range 10$^{35}$--10$^{38}$ ergs cm$^{-2}$ s$^{-1}$ is shown in
Table~\ref{tab:spin_l} with the associated source distance for each
luminosity.  If Roche lobe overflow is the source of accretion then a
luminosity of 10$^{38}$ ergs cm$^{-2}$ s$^{-1}$, close to the
Eddington luminosity, is expected \citep{hmxb_percent}.
Table~\ref{tab:spin_l} illustrates that this would locate the source
outside of the galaxy and would result in a clear spin-up of the
pulsar of $\sim$24 seconds between the BeppoSAX and XMM observations,
which is not seen.  The limit of $\dot{P}_{pulse}$$\sim$0.02 s
yr$^{-1}$, would imply that the source has a luminosity of 10$^{35}$
ergs cm$^{-2}$ s$^{-1}$ which is very much lower than expected from a
Roche lobe overflow system but would be typical of a wind accretion
system.  Hence, this is another indication that the donor star has not
filled its Roche lobe and that the pulsar is powered entirely by wind
fed accretion and thus lending further credence to the constraints on
the donor star given by the shaded region of Figure~\ref{mass_radius}. 

The soft excess ($<$ 3keV) evident in the XMM-EPIC data is identical
in the on and off pulse spectra implying that the soft excess may not
be pulsed.  Performing timing analysis on the events below 3 keV does
not show any period, however, the event rate is not high enough to
reach a definitive conclusion.  Soft X-ray excess have been found in
other HMXB sources, e.g. Vela X-1, and have been explained by
scattering or partial ionisation of the stellar wind.

The equivalent width of the 6.4 keV Fe line has a 3 sigma upper limit
of 40 eV and 25 eV for the off pulse and on pulse spectra
respectively.  Such values are consistent with $N_{H} <$ 5.0 $\times$
$10^{22}$ cm$^{-2}$ for a spherical geometry.  The non-detection of
the Fe line is therefore consistent with the absorbing column density
which is oberserved.

The spectral fit of the EPIC and ISGRI spectra indicate a column
density of $N_{H} =$ 6.8 $\pm$ 1.0 $\times$ $10^{22}$ cm$^{-2}$.  This is
much greater than the expected line of sight column
density,  $N_{H} =$ 1.0 $\times$ $10^{22}$ cm$^{-2}$, and suggests that the
absorption is intrinsic to the source and not the line of sight.  This
could be explained by the stellar wind from the OB supergiant
accreting onto the neutron star forming a dense spherical shell.  The
low statistics of the BeppoSAX data used by {\it Au03} did not allow
an accurate spectral analysis.  They inferred a luminosity of
$\sim$5.6 $\times$ 10$^{35}$ ergs s$^{-1}$ over the 1.8-10 keV range
by fitting their data to a power law with $\Gamma$ = -0.1 $\pm$ 0.1
and N$_{H}$ = 1.7$^{+0.8}_{-0.9}$ $\times$ $10^{22}$ cm$^{-2}$, and
assumed a source distance of 10 kpc.  The XMM spectral fit indicates
that the column density is much higher and hence the source luminosity
was previously underestimated.

Assuming that IGR J18027-2016 is 10
kpc away the 2-10 keV flux during the primary pulse of F$_{2-10 keV}$
= 8.9 $\times 10^{11}$ ergs cm$^{-2}$ s$^{-1}$ gives an estimate of
the source luminosity of $\sim$1.1 $\times 10^{36}$ ergs s$^{-1}$,
a typical luminosity for a non-Roche lobe filling Supergiant HMXB
\citep{MXB_integral}.  A useful
comparison is provided by 4U 1538-52, which is
a typical HMXB accreting from a stellar wind and shows similarities to
IGR J18027-2016.  4U 1538-52 has a long spin period of $\sim$529 s, an
orbital period of 3.75 days and undergoes a well defined X-ray
eclipse.  The companion is a 19.9$M_{\sun}$ B0 star \citep{4U_1538_comp} and the X-ray flux
has been estimated as $\sim$2 $\times$ 10$^{36}$ ergs s$^{-1}$
for a distance of 6.4 kpc \citep{4U_1538_flux}.  Hence the luminosity
of IGR J18027-2016 appears to be typical of wind accreting HMXBs
assuming it has a distance which approximates to the far side of the
Galactic Centre.

\section{Summary}

We presented an analysis of INTEGRAL, BeppoSAX and XMM-Newton
observations of IGR J18027-2016.  We conclude that the source is an
eclipsing HMXB system comprising an X-ray pulsar accreting matter
from the stellar wind of a late O -- early B supergiant.  The source
is persistent and has a high intrinsic
photoelectric absorption and is similar in this nature to a number of
other new INTEGRAL sources \citep{roland}.  IGR
J18027-2016 has the properties of a typical wind accreting HMXB system.

\begin{acknowledgements}
      This work is based on observations obtained with INTEGRAL and
      XMM-Newton, two ESA science missions with instruments, science
      data centre and contributions funded by the ESA member states
      with the participation of the Czeck Republic, Poland, Russia and
      the USA.

      We thank R. Cornelisse, S. Laycock and J. Zurita for their
      input and discussions.  A.~B. Hill acknowledges
      funding support from a PPARC PhD studentship.  We thank the
      anonymous referee for their useful comments.
\end{acknowledgements}


\begin{thebibliography}{}

\bibitem[{{Augello} {et~al.}(2003)}]{sax}
{Augello}, G., {et~al.}, 2003, ApJ, 596, L63

\bibitem[{{Bird} {et~al.}(2004)}]{ibissurvey}{Bird}, A.~J., {Barlow}, E.~J., {Bassani}, L., {et~al.}, 2004, ApJ, 607L, 33B 

\bibitem[{{Bowers \& Deeming}(1984)}]{roche_rad}{Bowers}, R. \&
  Deeming, T., 1984, Astrophysics I: Stars, Jones \& Bartlett, 319 

\bibitem[{{Corbet}(1986)}]{Corbet}{Corbet}, R.~H.~D., 1986, MNRAS, 220, 1047 

\bibitem[{{Frank} {et~al.}(1992)}]{spin_up}{Frank}, J.,
  {King}, A.~R., \& {Raine}, D.~J., 1992, in Accretion Power in Astrophysics, (Cambridge University
  Press), p. 125

\bibitem[{{Gros} {et~al.}(2003)}]{psle}{Gros}, A., {Goldwurm}, A., {Cadolle-Bel}, M., {Goldoni}, P., {Rodriguez}, J., {Foschini}, L., {Del Santo}, M., \& {Blay}, P., 2003, A\&A, 411, L179

\bibitem[{{Herrero} {et~al.}(1992)}]{OB2}{Herrero}, A., {Kudritzki}, R.~P., {Vilchez}, J.~M., {et~al.}, 1992, A\&A, 261, 209

\bibitem[{{Horne \& Baliunas}(1986)}]{period_err}
{Horne}, J.~H., \& {Baliunas}, S.~L., 1986, ApJ, 302, 757

\bibitem[{{Kawano \& Higuchi}(1995)}]{bootstrap}
{Kawano}, H., \& {Higuchi}, T., 1995, GeoRL, 22, 307K

\bibitem[{{Kaper} {et~al.}(2004)}]{hmxb_percent}{Kaper}, L., {Van der Meer}, A., \& {Tijani}, A.~H., 2004, RMxAC, 21, 128K

\bibitem[{{Lebrun} {et~al.}(2003)}]{isgri}{Lebrun}, F., {et~al.}, 2003, A\&A, 411, L141

\bibitem[{{Liu} {et~al.}(2000)}]{HMXB_Liu}{Liu}, Q.~Z., {Paradijs}, J.~van, \& {Heuvel}, E.~P.~J.~van~den, 2000, A\&As, 147, 25

\bibitem[{{Lutovinov} {et~al.}(2005)}]{luto_gc}{Lutovinov}, A., {et~al.}, 2005, A\&A, 430, 997

\bibitem[{{Nagase}(1989)}]{nagase}
{Nagase}, F., 1989, PASJ, 41, 1

\bibitem[{{Negueruela \& Coe}(2002)}]{HMXB_Coe}{Negueruela}, I. \& {Coe}, M.~J., 2002, A\&A, 385, 517

\bibitem[{{Negueruela}(2004)}]{MXB_integral}{Negueruela}, I., 2004,
  The Many Scales of the Universe - JENAM 2004 Astrophysics Reviews,
  Kluwer Academic Publishers. (astro-ph/0411759)

\bibitem[{{Press \& Rybicki}(1989)}]{fast_lsp}{Press}, W.~H. \& {Rybicki}, G.~B., 1989, ApJ, 338, 277P

\bibitem[{{Revnivtsev} {et~al.}(2004)}]{russian1}
{Revnivtsev}, M., {et~al.}, 2004, AstL, 30, 382R

\bibitem[{{Reynolds} {et~al.}(1992)}]{4U_1538_comp}{Reynolds}, A.~P.,
  {Bell}, S.~A., \& {Hilditch}, R.~W., 1992, MNRAS, 256, 631

\bibitem[{{Robba} {et~al.}(2001)}]{4U_1538_flux}{Robba}, N.~R., {Burderi}, L., {Di Salvo}, T., {Iaria}, R., {Cusumno}, G., 2001, ApJ, 562, 950

\bibitem[{{Scargle}(1982)}]{scargle}
{Scargle}, J.~D., 1982, ApJ, 263, 835 

\bibitem[{{Str\"{u}der} {et~al.}(2001)}]{xmm-epic}{Str\"{u}der}, L., {Briel}, U., {Dennerl}, K., {et~al.}, 2001, A\&A, 365, L18

\bibitem[{{Turner} {et~al.}(2001)}]{xmm-mos}{Turner}, M.~J.~L., {Abbey}, A., {Arnaud}, M., {et~al.}, 2001, A\&A, 365, L27

\bibitem[{{Ubertini} {et~al.}(2003)}]{ibis_fov}{Ubertini}, P., {et~al.}, 2003, A\&A, 411, L427

\bibitem[{{Walter} {et~al.}(2003)}]{roland}{Walter}, R.,
  {Courvoisier}, T.~J.~-L., {Foschini}, L., {et~al.}, 2004, ESA SP
  552: Proceedings of the 5th INTEGRAL workshop - The INTEGRAL
  Universe, Editors: V. Sch\"{o}nfelder, G. Lichti \& C. Winkler, p417

\bibitem[{{White} {et~al.}(1995)}]{pulsar_spec}{White}, N.~E.,
  {Nagase}, F., \& {Parmar}, A.~N., 1995, in X-Ray Binaries, eds W.~H.~G. Lewin,
  J. van Paradijs \& E.~P.~J. Van den Heuvel (Cambridge University
  Press), p. 32

\bibitem[{{Wilson \& Dopita}(1985)}]{OB1}{Wilson}, I.~R.~G. \& {Dopita}, M.~A., 1985, A\&A, 149, 295

\bibitem[{{Winkler} {et~al.}(2003)}]{integral}{Winkler}, C., {et~al.}, 2003, A\&A, 411, L1

\end{thebibliography}
\end{document}